# Dude, Where's My (Autonomous) Car? Defining an Accessible Description Logic for Blind and Low Vision Travelers Using Autonomous Vehicles

Paul D. S. Fink[1,2]; 0000-0003-2915-1331
Justin R. Brown[1]; 0000-0002-3359-8411
Rachel Coombs[1]; 0009-0004-8593-559X
Emily A. Hamby[1]; 0009-0005-0318-1446
Kyle J. James[1]; 0009-0003-3205-824X
Aisha Harris[1]; 0009-0008-0916-0912
Jacob Bond[3]; 0000-0003-2025-5230
Morgan E. Andrulis[3]; 0009-0009-8827-4885
Nicholas A. Giudice[1,4]*; 0000-0002-7640-0428

[1]VEMI Lab, The University of Maine, Orono, Maine 04469, USA.
[2]College of Computing, Grand Valley State University, Allendale, Michigan 49401, USA.
[3]Connected Vehicle Experience Research Lab, General Motors Global Research & Development, Warren, Michigan 48092, USA.
[4]School of Computing and Information Science, The University of Maine, Orono, Maine 04469, USA.

*Corresponding Author: nicholas.giudice@maine.edu

**Purpose:** Autonomous vehicles (AVs) are becoming a promising transportation solution for blind and low-vision (BLV) travelers, offering the potential for greater independent mobility. This paper explores the information needs of BLV users across multiple steps of the transportation journey, including finding and navigating to, entering, and exiting vehicles independently.

**Methods:** A survey with 202 BLV respondents and interviews with 12 BLV individuals revealed the perspectives of BLV end-users and informed the sequencing of natural language information required for successful travel. Whereas the survey identified key information needs across the three trip segments, the interviews helped prioritize how that information should be presented in a sequence of accessible descriptions to travelers.

**Results:** Taken together, the survey and interviews reveal that BLV users prioritize knowing the vehicle's make and model and how to find the correct vehicle during the navigation phase. They also emphasize the importance of confirmations about the vehicle's destination and onboard safety features upon entering the vehicle. While exiting, BLV users value information about hazards and obstacles, as well as knowing which side of the vehicle to exit. Furthermore, results highlight that BLV travelers desire using their own smartphone devices when receiving information from AVs and prefer audio-based interaction.

**Conclusion:** The findings from this research contribute a structured framework for delivering trip-related information to BLV users, useful for designers incorporating natural language descriptions tailored to each travel segment. This work offers important contributions for sequencing transportation-related descriptions throughout the AV journey, ultimately enhancing the mobility and independence of BLV individuals.

**Statements and Declarations**

**Funding:** This work was supported by General Motors Global Research & Development. **Competing Interests:** The authors have no competing interests to declare that are relevant to the content of this article. **Human Subjects Approval:** This research was approved by The University of Maine Institutional Review Board under application #2023-08-09. Informed consent was obtained from all participants included in this research.

## 1 Introduction

Transportation is a challenging undertaking for the 1.3 billion people experiencing significant disabilities worldwide [15], including the approximately 295 million people who report moderate to severe visual impairment [3]. When considering 'transportation', a common misconception is that the travel process begins and ends in the vehicle. In reality, the complete journey is characterized by several mutually dependent travel segments that start with planning the trip. Among the remaining segments are navigating to a ride, entering, exiting, and safely navigating to the destination. Autonomous vehicles (AVs) hold the potential to revolutionize accessible mobility for blind and low-vision (BLV) travelers by providing door-to-door transportation without requiring friends, family members, or sighted guides for assistance. However, this promise of independent mobility depends on future interfaces conveying meaningful and useful information across each segment of the complete trip. While the body of work investigating accessible AVs for BLV users has prioritized in-vehicle interaction, much less is known about information needs in the equally important (but often ignored) segments that involve localizing a vehicle, entering it safely, and exiting and orienting to the destination [20, 30].

To address these problems, we sought to identify what we refer to as an *accessible description logic*, providing inclusively designed information for BLV travelers in a user-defined sequence of natural language descriptions across otherwise disconnected travel segments. Our description logic aims to be a progression of information categories that designers and developers can use to insert natural language audio cues that are adapted to the specific environment. Just as a developer might use a sequence of variables to 'slot in' specific values in that sequence, we envision the utility of this logic to be identifying the order and priority of information that can be adapted to each AV trip.

We specifically study aspects of AV-enabled transportation known to be difficult for BLV people, starting first with safely navigating to and localizing a ride once it arrives. As AVs adopting rideshare models may arrive to unanticipated locations in dynamic environments [17, 22], BLV travelers will likely be required to utilize outdoor navigation skills to find and navigate to AVs, often in tandem with applications designed to support wayfinding [22], in addition to their preferred mobility tool, e.g., the long cane or guide dog. Unlike typical outdoor navigation activities, however, localizing AVs involves a host of context-specific tasks for BLV travelers, including identifying and confirming the correct vehicle, avoiding obstacles within the pick-up area (e.g., construction hazards), and finding the correct door handle [20]. While the tasks for navigating to a vehicle are well-defined in the literature, how to prioritize and order information in an accessible interface remains an open question. This is also the case for entering and orienting within the vehicle, which often includes determining how to enter safely and understanding seat locations and availability. Finally, once the AV arrives at its destination, BLV users must exit the vehicle safely and orient themselves to another environment [30]. Rather than studying each of these segments in isolation, as is common in the existing literature, we seek to provide users and designers with a comprehensive and consistent logic for completing the trip. Our goal in this paper is to identify a structured sequence of natural language descriptions for each of these travel segments, ensuring that user needs are reflected throughout by directly engaging BLV users in two studies. The first, a confidential, anonymous survey with ($N$ = 202) BLV respondents, was motivated by the need to identify essential information to be conveyed during the trip. It sought to answer our first research question:

**RQ1: What information is important for BLV travelers to receive across the complete trip with AVs?**

The second stage of this project consisted of interviews with ($N$ = 12) BLV individuals, which aimed to order the initial set of information preferences defined by the survey in a logical sequence for each trip segment. This goal was addressed by our second research question:

**RQ2: How should information be ordered and presented to BLV travelers across travel segments with AVs?**

The interviews provided feedback not only on the order in which participants preferred to receive information, but also the accessibility aids and technologies they found most useful, and enabled a clearer understanding of their individual transportation experiences. By integrating survey results and interview insights, we developed a final description logic and accompanying guidelines that offer best practices for designing accessible transportation apps and related technologies. The key contributions of this work advance accessible computing in the transportation context by providing a structured approach to delivering critical trip-related information for BLV users, something that to our knowledge has not been previously studied. Importantly, this project addresses a critical user experience gap (e.g., BLV riders) identified by our OEM partner, while also providing a strong foundation for future studies that empirically test the logic defined here using behavioral methods. We believe that our findings help pave the way for ongoing work by our group and others addressing the question at the end of the paper (Section 6.3) regarding the efficacy of using audio-only vs. multisensory information presentation in user interfaces (UIs) supporting AV usage as part of a complete trip solution. We highlight the apparent experiential gap indicated by BLV users who self-report preferences for audio-only information when, in related work, evidence suggests performance is on-par or better with multisensory information. By identifying audio-only information in the AV context across multiple trip segments, the current paper contributes an important testbed for evaluating this potential gap between what people say they prefer (which we hypothesize is based on their existing experiences with audio-only interfaces) and how they perform with new multisensory user interfaces.

The remainder of this paper is organized as follows: In Section 2, we review the current work in AV technology and its impact on independent transportation for BLV individuals, including research on complete trip accessibility. Section 3 details the methodology and results of Study 1, which identifies critical trip information. Section 4 presents the methods and results of Study 2, which refines how that information should be structured and delivered as well as reports the findings of thematic analysis on the interviews as a whole. In Section 5, we present our culminating description logic and in Section 6 we discuss our findings in relation to existing accessibility literature and AV design. Finally, Section 7 concludes with the key takeaways from this work.

## 2 Related Work

AV technology is rapidly advancing; however, the small but growing body of literature examining accessible AVs for BLV users indicates that more work is needed to meet the needs of travelers who often rely on human assistance throughout the trip [30]. One useful strategy in pursuit of this goal is involving end-users in the design of AV systems. Indeed, multiple researchers have emphasized the importance of the disability community being involved in the development of technologies that are needed to make AVs readily accessible to consumers. For instance, in their focus group study, Brinkley et al. (2017) found that BLV users were concerned about not being adequately considered in self-driving development [11], industry white papers have elucidated the important role of advocacy among people with disabilities to be considered in AV development [13], and user experience work has emphasized inclusive, participatory design of AV system features to support BLV individuals [20]. While this direct involvement is important, Ma (2024) correctly points out that prior work is predominantly hypothetical, as AVs truly designed for BLV users do not yet exist [28]. Recognizing this limitation, the following subsections review the available body of work exploring BLV AV accessibility, as well as the importance of natural language descriptions in supporting the complete trip, which we argue is critical to consider as travel is not limited to discrete trip segments or activities.

### 2.1 Complete Trip Accessibility for BLV Travelers

The emergence of AVs presents both opportunities and challenges for BLV travelers. While AVs hold the potential to enhance mobility, independence, and workforce participation [21], research consistently highlights the need for concerted inclusive design efforts to address accessibility concerns. For instance, in their survey-based study, Bennett et al., 2020 noted that BLV users are skeptical that AVs will be designed for them [2], which was a concern echoed by participants in a dual survey and focus group study by Brinkley and colleagues [9]. Even understanding these concerns, the previous work emphasized optimism about AVs among the BLV community, which has been demonstrated to be more prevalent among BLV users than their sighted counterparts [24]. One key issue is ensuring that BLV persons are effectively supported throughout the entire travel experience, from planning a trip to reaching their destination. Studies have shown that BLV travelers rely on various cues to understand the travel environment across the trip, many of which are currently provided by human rideshare drivers [6, 8]. For instance, Brewer and Ellison (2020) found that BLV rideshare users often depend on drivers for assistance in locating the vehicle, exiting safely at accessible points, and receiving verbal descriptions of the environment [5]. The predicted transition to AV rideshare services raises concerns about whether these supportive interactions currently provided by humans can be effectively replaced by technology when human drivers are no longer in the loop [7, 18]. While past rideshare experiences based on human-driven vehicles can highlight current accessibility gaps, they do not directly translate to a one-to-one set of expectations or solutions for AVs, when the human agent is no longer part of the information-access equation. Future AV use will likely involve new interaction paradigms, different constraints, and evolving user priorities, necessitating distinct design strategies. Other research has addressed the lack of support provided to BLV users from modern navigation applications and devices that do not meet users' requirements, such as different types of intersections all being labeled as 'intersection' and implementations that do not accurately address the precise location and orientation of crosswalks [16, 37]. Non-visual interfaces have been explored to bridge this gap, primarily focusing on in-vehicle experiences providing information on vehicle location via in-air haptics and/or spatialized audio [19], as well as on-the-go updates about the ride via audio cues [10]. Recent work has also explored the ideal combination of multisensory cues for BLV users during the in-vehicle portion of the route, identifying tactile and spatialized audio as the optimal combination [12]. However, less attention has been paid to other phases of the journey like pre-trip planning and vehicle localization [17, 34]. While some work has focused on designing inclusive interfaces for localizing AVs [20] and exiting AVs [30], no research to date has explored a unified and accessible framework across multiple stages of the trip for BLV travelers. Moreover, no work has sought to identify how to appropriately structure information sequentially for users as they navigate these multiple trip segments. Addressing both of these issues represent key contributions of the current work. A promising approach is to leverage users' interest in utilizing speech and audio interaction with AVs [11, 18], as explored in the following.

## 2.2   Natural Language Sequences for Spatial Tasks

Turn-by-turn (TBT) direction systems have become the de facto standard for audio-based navigation, but evidence suggests that TBT natural language systems can be problematic for BLV users, as the information provided is often inadequate for developing an accurate mental representation necessary for safe navigation [23]. This is primarily because of the information (or lack thereof) provided in standard TBT direction systems, leading to significant errors [1]. For BLV users to safely and accurately travel to and from AVs, information beyond the route is necessary as well, including indication of relevant landmarks, obstructions, cues about sidewalk or street parameters, and car orientation [20]. An alternative to traditional TBT direction systems for BLV travelers are systems that include spatialized information, i.e., where the distance and direction of a street, landmark, or point of interest is audibly received from its location in space [38]. While adding spatialized information to TBT instructions can improve route guidance accuracy and reduce cognitive load [26, 27], it does not eliminate the limitation that these instructions are only predominately conveying route guidance information. Therefore, to address this limitation and offer a more comprehensive navigation experience, this study explores the potential of enhancing NL navigation instructions as an alternative to traditional TBT directions. Moreover, we aim to understand how to best sequence this information to aid future navigation systems and the associated algorithms that provide users with sequential real-time information [35].

## 3   Study 1: Survey of BLV Information Needs Across the Complete Trip

To determine the information needs for BLV travelers across the vehicle localization, entry, and egress segments of the trip (RQ1), a survey was developed by the authors of this paper and distributed remotely to ($N$ = 202) BLV participants by Qualtrics (https://www.qualtrics.com/). The decision to use a third-party survey service for participant recruitment was intended to maximize respondents who met demographic criteria of interest (see 3.1.1 Participants below). Prior to providing responses, participants were asked to imagine three scenarios that involved finding, entering, and exiting an AV in an urban area. To increase the realism of the imagined scenarios, participants were asked to imagine that the AV would be taking them to a grocery store. The AV was noted as being able to safely and legally provide transport without being accompanied by a human driver or attendant. To reflect the possibility of widespread mobility-as-a-service prediction models for future AV deployment [14, 31], participants were also told that there might be other passengers on board and that the seats might be in nontraditional arrangements (like a bus or a shuttle). Due to the limited real-world availability of AV transportation, the use of these 'imaginative' scenarios were employed that aimed to unify participant understanding and encourage responses *as if* they were traveling in these environments.

**3.1 Methods**

The survey consisted of 32 questions (Appendix A) divided between three parts of the trip: navigating to the vehicle, entering and orienting within it, and exiting and orienting to the surrounding environment. Survey respondents were instructed to imagine that they were going to use an autonomous rideshare vehicle in three scenarios related to each part of the trip. Individual pieces of information that were evaluated within these three phases were generated based on the prior literature on entering and exiting AVs [20, 30], as well as from personal insight and lived experience of one of the authors on this paper who is congenitally blind and a frequent traveler. For example, in the navigation phase, information included finding the correct vehicle and door handle, as well as avoiding obstacles. In the entering phase, eight questions evaluated a range of information items (e.g., seat placement, cleanliness details, information on other passengers). The exiting phase also included eight questions evaluating a range of information items (e.g., obstacle avoidance, other passenger movement, and points of interest). These information items were first rated on 5-point Likert scales to evaluate their importance (1, Strongly Disagree to 5, Strongly Agree). To gain an initial understanding of how to order information in our description logic (RQ2), participants were then asked to identify which piece of information they would like to receive first during that phase. The survey was tested internally for accessibility and confirmed to work well with screen readers prior to deployment. The research was approved by The University of Maine IRB.

*3.1.1 Participants*

Selection criteria ensured participants were at least 18 years of age with a known and uncorrected visual impairment, utilized an accessibility device or mobility aid (e.g., screen reader, magnification, white cane, guide dog, etc.), and were non-drivers who utilize transportation, such as rideshares, public transit, or private vehicles. Demographic information such as name, age, gender, and email addresses were not able to be recorded by Qualtrics; however, geographic level of urbanization, the classification of vision status, and what accessibility aids participants utilize were all documented. The majority of participants lived in suburban areas ($N = 81$), classified their vision status as "Moderate Vision Impairment" ($N = 93$), and reported using magnification ($N = 133$). A complete summary of participant demographics from the pre-survey questions are detailed in Table 1. Participants were compensated by Qualtrics in accordance with their compensation protocols.

Table 1 Interview Participant Self-Reported Demographic Information

| Geographic Area | | Vision Status | | Accessibility Aids (multi-select) | |
|---|---|---|---|---|---|
| Category | N (%) | Category | N (%) | Category | N (%) |
| Suburban | 81 (40.1%) | Moderate vision impairment | 93 (46%) | Magnification | 133 (65.8%) |
| Urban | 72 (35.6%) | Mild vision impairment | 61 (30.2%) | Mobile screen readers | 93 (46%) |
| Rural | 49 (24.3%) | Severe vision impairment (includes blindness) | 48 (23.8%) | PC screen readers | 49 (24.3%) |
| | | | | White cane | 36 (17.8%) |
| | | | | Guide dog | 32 (15.8%) |
| | | | | CV apps | 30 (14.9%) |
| | | | | Braille display | 13 (6.4%) |
| | | | | Other* | 11 (5.4%) |

*3.1.2 Data Analysis*

The Likert scale questions (1, Strongly Disagree to 5, Strongly Agree) rating the importance of each item in the survey were first analyzed using basic descriptive statistics; mean (*M*) and standard deviation (*SD*). As our goal was to identify whether each item was important, rather than to compare the importance between the individual items, inferential statistics were not employed for the results of these questions. However, subsequent inferential analyses of the distribution of rankings by vision impairment status (Mild, Moderate, and Severe) were conducted to reveal potential differences or similarities in item rankings among these groups. As the rankings are non-parametric data and three visual groups are being compared, a Kruskal-Wallis test was used to determine the impact of vision status for each item ranking. Dunn's post-hoc comparisons with Bonferroni correction were conducted on items where the Kruskal-Wallis test was found to be significant. The results and Likert distributions, separated by survey section, are reported below in sections 3.2.1, 3.2.2, and 3.2.3.

## 3.2 Results

*3.2.1 Navigation*

Based on the self-reported Likert data, the majority of participants wanted information about the make, model, and year of the vehicle ($M = 4.15$, $SD = 1.05$). Participants also indicated that they found it challenging while navigating to the vehicle to avoid hazards ($M = 3.53$, $SD = 1.20$) and to find the correct vehicle ($M = 3.51$, $SD = 1.23$). Participants did not respond as strongly for or against it being challenging for them to locate the correct door and handle when navigating to the vehicle ($M = 3.14$, $SD = 1.32$). Figure 1 displays the total distribution of rankings and the distributions by reported vision status. Three of the items (Q1, Q2, Q3) were found to have significant differences in rankings between vision status groups ($p < .001$). Post-hoc comparisons revealed significant differences in rankings (all $p$'s $< .01$) between the Mild and Severe groups as well as the Moderate and Severe groups across the three items. These results suggest that individuals with severe vision impairment may face greater challenges in locating the correct vehicle, identifying the correct door, and avoiding hazards on the way to the vehicle compared to those with mild or moderate vision impairment.

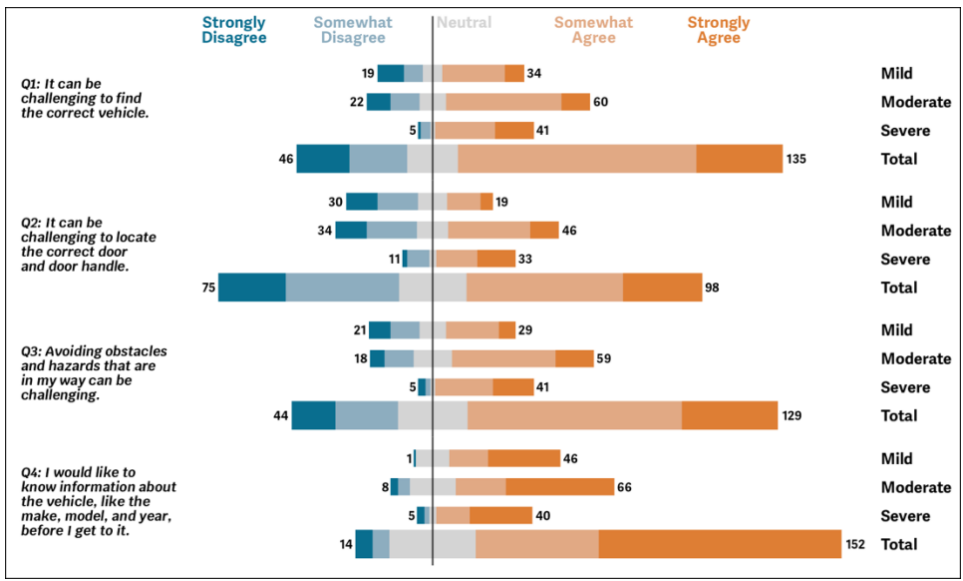

**Fig. 1** Diverging stacked bar plots showing the distribution of Likert responses by self-reported vision status for Study 1 Survey statements concerning Navigating to the Vehicle

When considering the piece of information people want to know first (Figure 2), several information items received roughly 40 responses (~20% of participants). These included knowing which door to enter, the direction of the vehicle, position of other passengers, information about the curb, and information about the destination.

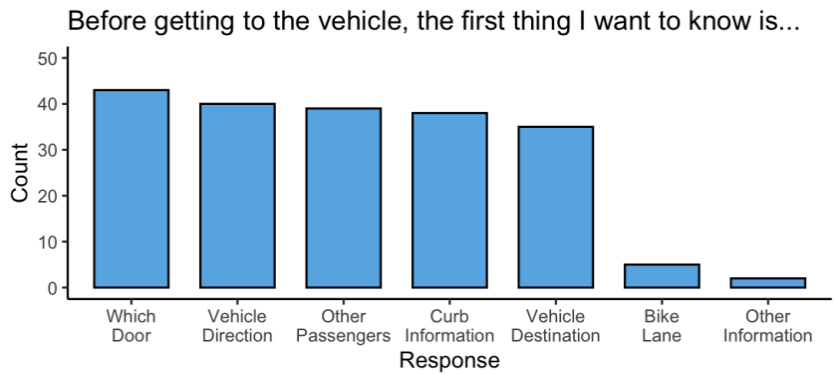

**Fig. 2** Bar plot of Survey 1 responses about the first information participants wanted during the Navigating to the Vehicle phase

To determine if there were statistically significant differences between the items listed first, a Chi-square goodness of fit test was performed ($\chi^2 = 63.72$, $p < .001$), indicating that participant preferences were not equally distributed across the range of information types (i.e., the Response categories in Figure 2). That is, results suggest that there was an overall difference between the information to be delivered first during the navigation phase. This was likely due to the few responses in the *Bike Lane* and *Other Information* categories. Indeed, pairwise post-hoc tests demonstrated that *Other Information* and *Bike Lane* were chosen statistically significantly fewer times than the other information categories (all $p$'s < .001). This was surprising since the *Other Information* category was included to capture responses like the make, model, and year of the vehicle, which participants indicated they wanted strongly in the earlier Likert questions (Q4 from Figure 1). Furthermore, the post-hoc tests demonstrated no statistically significant differences between knowing which door to enter, the direction of the vehicle, position of other passengers, information about curbs, and information about the destination. Taken together, these data indicate that a range of information is important to BLV users when navigating to AVs and that users are divided on which information should be presented first (which we analyze further in the Study 2 interviews, Section 4).

*3.2.2 Entering and Orienting*

Across all Likert scale questions about entering the vehicle (Figure 3), participants responded that they want to know each piece of information queried in the survey, with all means except the location of the control interface ($M = 3.93$, $SD = .10$) scoring a 4 or above. Two of the nine items, the vehicle's route (Q2) and interior seat layout (Q4), were found to have significant differences between vision status groups ($p < .05$). Post-hoc comparisons revealed significant differences between the Moderate ($M = 4.05$, $SD = .94$) and Severe ($M = 4.38$, $SD = 0.94$) groups for Q2 rankings ($p = 0.04$) and significant differences between Mild ($M = 4.02$, $SD = .90$) and Severe ($M = 4.48$, $SD = .85$) groups for Q4 rankings ($p < .01$). These results may suggest that participants with more severe vision impairment prioritize certain vehicle entry information more strongly – specifically, clarity about the vehicle's planned route and how seats are arranged – than those with mild or moderate vision impairment.

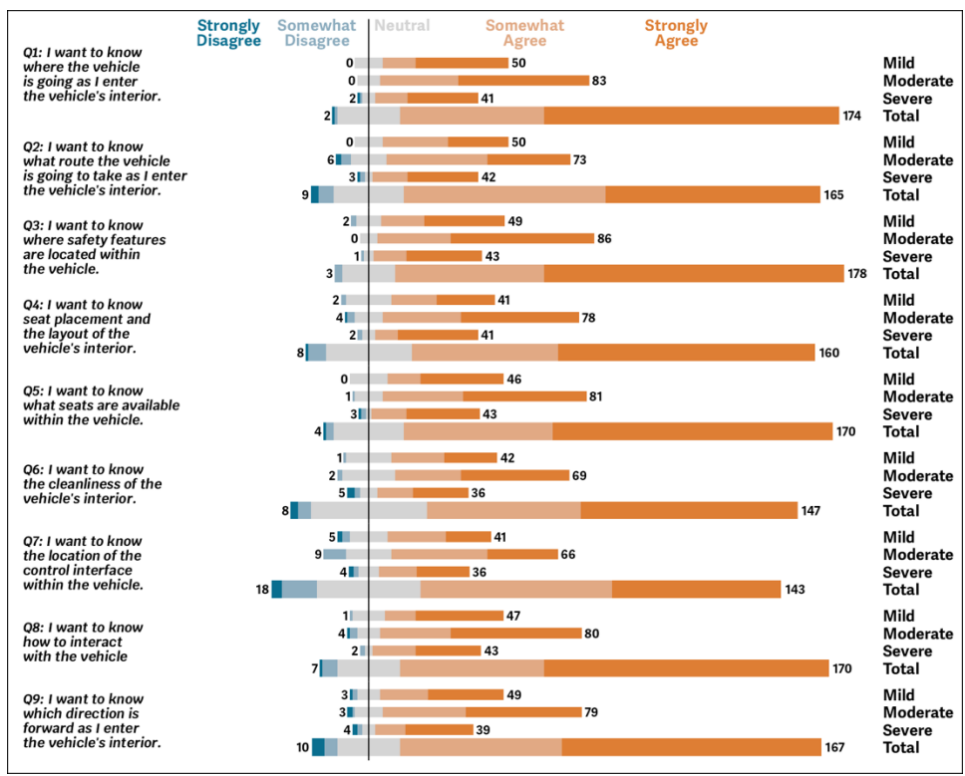

**Fig. 3** Diverging stacked bar plots showing the distribution of Likert responses by self-reported vision status for Study 1 Survey statements concerning Entering and Orienting

When asked what information participants wanted to know first in the entering phase (Figure 4), the most common response was information about the vehicle's destination with 46 responses, or 22.8% of participants, closely followed by the location of safety features ($N = 43$, 21.3%).

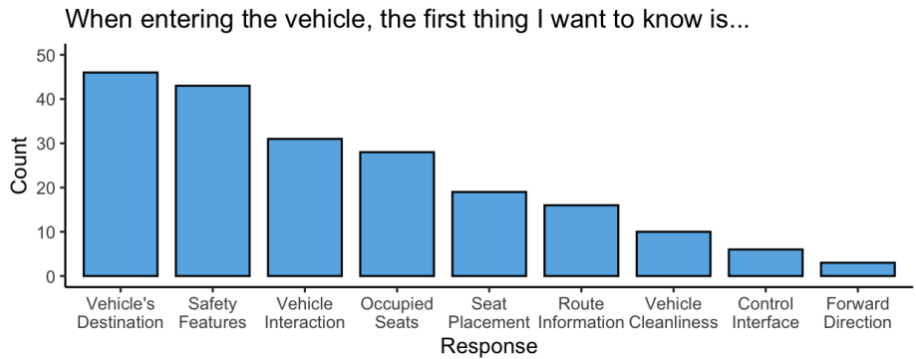

**Fig. 4** Bar plot of Survey 1 responses about the first information participants wanted during the Entering and Orienting phase

A Chi-square goodness of fit test was performed ($\chi^2 = 86.35$, $p < .001$), indicating that the information items were not equally distributed across information types. Pairwise post-hoc tests demonstrated that the *Vehicle's Destination* and *Safety Features* were statistically more likely than the five least common responses (*Seat Placement, Route Information, Vehicle Cleanliness, Control Interface,* and *Forward Direction*), with all *p*'s less than .05. This indicates that during the entry phase of the trip, BLV travelers are concerned with verifying the vehicle's ultimate destination prior to entry and to know how to travel safely.

*3.2.3 Exiting and Orienting*

The results of the Likert scale questions in the exiting and orienting section of the survey (Figure 5) indicate that users strongly desire access to a range of information when exiting the vehicle, with all means above 4.24, with the exception of points of interest in the immediate area ($M = 3.86$, $SD = 1.09$). Two of the eight items, where the vehicle is located when exiting (Q7) and the direction passengers are facing when exiting (Q8), were found to have significant differences between vision groups ($p = .05$). Post-hoc comparisons revealed significant differences between Mild ($M = 4.13$, $SD = 1.04$) and Moderate ($M = 4.52$, $SD = .75$) groups for Q7 ($p = .05$) and significant differences between Mild ($M = 4.05$, $SD = .92$) and Moderate ($M = 4.43$, $SD = .70$) groups for Q8 ($p = .03$), suggesting that individuals with moderate vision impairment may place greater importance on knowing the vehicle's exact location and their orientation upon exiting the vehicle compared to those with mild impairment.

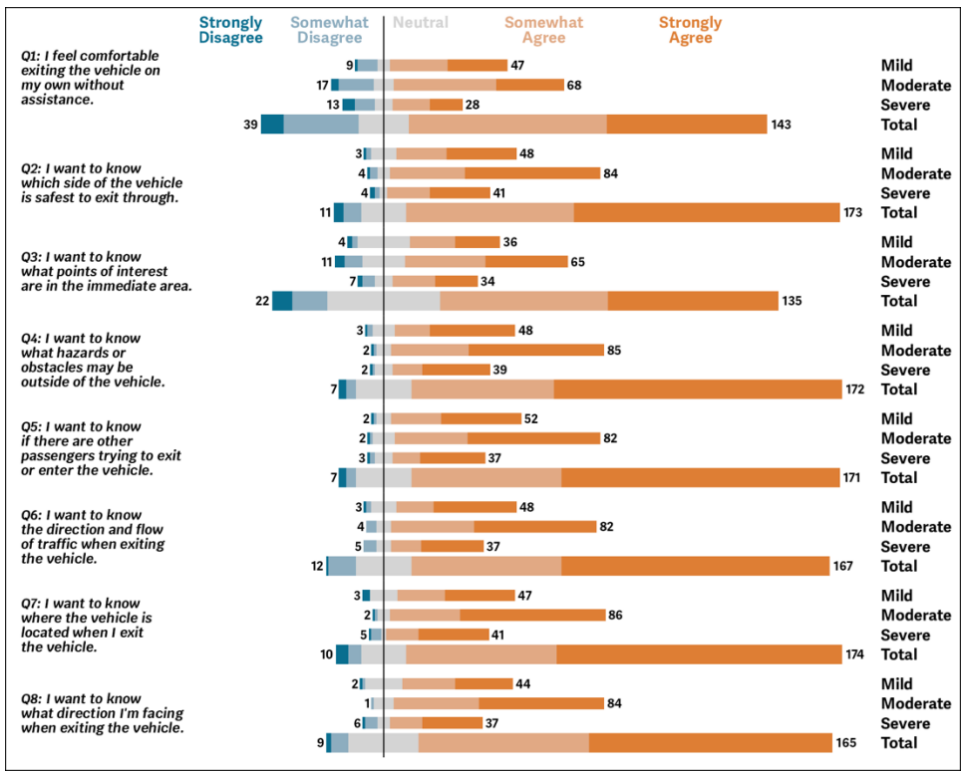

**Fig. 5** Diverging stacked bar plots showing the distribution of Likert responses by self-reported vision status for Study 1 Survey statements concerning Exiting and Orienting

When asked what information about exiting they wanted to know first when immediately leaving the vehicle, 79 participants, or 39.1%, chose information about what hazards and obstacles may be outside the vehicle. Other participant responses to this question can be found in Figure 6, with the next most common response being which side of the vehicle they are exiting, indicated by 40 responses, or 19.8% of participants.

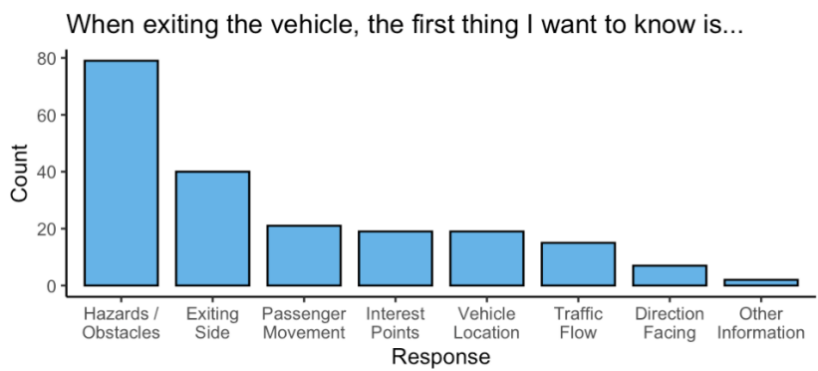

**Fig. 6** Bar plot of Survey 1 responses about the first information participants wanted during the Entering and Orienting phase

A Chi-square goodness of fit test was performed ($\chi^2 = 165.6$, $p < .001$), indicating that the information items were not equally distributed across information types. Pairwise post-hoc tests demonstrated that the *Hazards/Obstacles* item was statistically more likely than the other information types, with all $p$'s less than .001. *Exiting Side* was also demonstrated to be more likely than the six items with fewer responses, with all $p$'s less than .05. Taken together, these results again suggest a clear preference for safety through notification of hazards and obstacles, as well as which side of the vehicle is safe to exit.

## 4 Study 2: Interviews Identifying Desired Order of Description Information

After identifying the types of information that were important in Study 1, we sought to glean additional information on what, when, and how information should be presented to BLV travelers during an autonomous vehicle trip (RQ2). Whereas in Study 1 we queried participants on what information is important, as well as what information they would want *first* in three stages of AV travel, the goal in this subsequent interview study was to confirm the first piece of information (identified in the survey) and to prioritize each piece of information thereafter to inform our description logic. We engaged ($N$ = 12) BLV participants in an interactive interview that was guided by a pre-interview worksheet that participants completed in advance of the interview (see Section 4.1.2 Procedure).

### 4.1 Methods

*4.1.1 Participants*

A total of 12 BLV participants aged 31-71 ($M$ = 47.58, $SD$ = 15.55) who identified as legally blind and used a primary mobility aid were recruited across a range of visual impairment and etiology (Table 2). The interviews were conducted one-on-one with a researcher, with the participants being recruited through the lab's local network in the United States and via personal contacts in the BLV community. Given the nature of recruitment, no interviewees also completed the Study 1 survey. Participants were compensated with a $50 gift card.

Table 2 Interview Participant Self-Reported Demographic Information

| Participant | Age | Gender | Visual Condition | Visual Function | Mobility / Accessibility Aids |
|---|---|---|---|---|---|
| 1 | 71 | Male | Retinopathy of Prematurity | No Vision | White cane, Guide dog |
| 2 | 63 | Female | Medical Mishap | Movement, Light | White cane, Guide dog |
| 3 | 71 | Male | Hereditary Optic Neuropathy | Shapes, Shadows | White cane |
| 4 | 32 | Male | Retinitis Pigmentosa | Light, Colors | White cane |
| 5 | 31 | Male | Unknown | Light | White cane |
| 6 | 32 | Female | Leber's Congenital Amaurosis | Light | White cane |
| 7 | 34 | Female | Retinopathy of Prematurity, Glaucoma | Light, Colors | White cane, Guide dog, Braille |
| 8 | 56 | Male | Unknown | Light | Guide dog |
| 9 | 44 | Non-Binary | Retinitis Pigmentosa | Light, Shapes | White cane, Guide dog |
| 10 | 32 | Male | Condition with High Eye Pressure | No Vision | White cane |
| 11 | 54 | Female | Optic Nerve Hypoplasia | Light, Colors | White cane, Magnification |
| 12 | 51 | Male | Leber's Congenital Amaurosis | No Vision | White cane |

*4.1.2 Procedure*

Participants completed a pre-interview worksheet (Appendix B) prior to the interview and ranked the order that they would desire for information to be presented in each of the three travel segments. The results of the Likert responses from the survey in Study 1 prompted the addition of obstacle and hazard information as an option in the ordering for the navigation phase on the pre-interview worksheet. The worksheet asked participants to think about what information would be most important for them to receive first and then to imagine how the information should flow thereafter. Again, this exercise employed three scenarios that included (1) navigating to an AV, (2) entering and orienting in the vehicle, and (3) exiting the vehicle and orienting to the surrounding environment. For each imagined scenario, participants ranked the order in which they would like to receive the information by placing a "1" next to the information they would most prefer to receive first and increasing integers for the pieces of information they would want to know thereafter. Participants were able to eliminate pieces of information they found unimportant by ranking them with a "0".

After participants completed and sent the pre-interview worksheet back to the team, remote one-on-one interviews were conducted via Zoom, which were audio recorded. The interviews lasted approximately an hour each and followed a general script that was tailored to each participant based on their pre-interview worksheet answers, allowing for elaboration of choices and reasoning. To begin, participants were asked questions about their demographic information and current experiences with technology and transportation as well as questions soliciting their overall thoughts on the rollout of fully autonomous vehicles. Next, participants were asked to explain their thought process as they filled out the worksheet including ordering of information and information they eliminated. To help contextualize and explain results from the Study 1 survey, participants were also provided with the most frequent response to the "first thing I want to know is" questions from the survey for each of the three travel segments and asked to postulate why the response was indicated and under what scenarios it would be most relevant. These questions were intended to help resolve any disagreement between the first item results from the survey and the entire sequence responses collected from the interviews. Given our interest in how information should be presented (RQ2), participants were also asked if information should be presented from their phone, the vehicle, or another device, as well as the presentation modality (i.e., haptics, audio, or combinations of both). To conclude the interviews, participants were asked about their safety concerns related to AVs and what problems with their current transportation experience they think autonomous vehicles could solve. The full interview guide can be found in Appendix C.

*4.1.3 Data Analysis*

Qualitative thematic analysis of the interview transcripts was conducted by two independent researchers using the Taguette qualitative data analysis tool [33]. Going through the interview transcripts in opposite orders to avoid fatigue effects, the researchers identified key ideas (codes) and recurring themes that participants mentioned in their responses. Afterwards, a team of five researchers reviewed all of the codes and their associated quotes to resolve disagreements. This process generated 83 unique codes that were further categorized by the team of researchers into subgroups and thematic groups [4]. The qualitative results, including four emerging themes from the thematic analysis process, are reported in Section 4.2.1.

As non-parametric data, the order of information in the navigation, entry, and exiting segments was determined by calculating the mean rank followed by a Friedman's test. Items that participants eliminated in their worksheet, were scored as last place plus 1. In the event of a tie, the information item with the fewest number of eliminations, meaning more participants chose to include that information, would receive a higher ranking. How this information should be presented (i.e., the presentation modality) was also recorded. The most common response to presentation for each information item was recorded in Tables 3–5 along with the mean order of the information items. The results of the information sequencing are reported below in Section 4.2.2.

**4.2 Results**

*4.2.1 Qualitative Results*

Qualitative thematic analysis of participant interviews revealed four key themes that highlight the complex relationships that BLV individuals have with their current transportation experiences and the potential for AVs to significantly reshape their independence and access to mobility. First, *safety* emerged as a critical consideration, encompassing both immediate physical risks and the broader reliability of AVs as an emerging technology. Second, participants detailed the *difficulties and challenges* they currently face with transportation, including personal, spatial, and technological barriers. Third, participants expressed significant optimism regarding the *opportunities* presented by AVs, envisioning increased independence, agency, and efficiency in their future mobility. Finally, interviews revealed the *adaptations* BLV individuals utilize to navigate these challenges, often mentioning the need to balance several navigation apps at once, manage their time, and rely on others, all of which can be frustrating and inefficient. These four themes collectively describe the impact that transportation has on the lives of BLV individuals and underscore the potential of AV technology to transform their transportation experiences. Put simply, **P3** offered, "*The ability to independently travel when blind is a big deal.*" The following subsections are organized around the four themes identified from the interviews and offer additional insights into their role in AV transportation.

*Safety*

Participants expressed a range of safety concerns focused on environmental hazards, unexpected events, and the reliability of both current rideshares and future AVs. During the navigation phase, many emphasized that AV interfaces must respond to unpredictable environments. **P1** illustrated this by noting, "*You're somewhat vulnerable when there's some unusual situation, especially a hole in the ground [...] Things like that can be very difficult.*" Entering the vehicle raised concerns about identifying the correct car and feeling secure. **P3** shared, "*It's very important that the autonomous car makes me comfortable that I'm in the right place and that it's safe to get in.*" Exiting also posed challenges, particularly when vehicles stop in unsafe locations. **P8** remarked, "*I definitely want to know which side of the vehicle to get out on because I don't want to walk into headlong traffic.*" Concerns about AVs' ability to handle unexpected incidents—like accidents or road obstructions—were frequent. "*My safety concerns would be an autonomous car's ability to deal with the unexpected*" said **P3**. Worries extended to being dropped off in chaotic or unfamiliar places: "*I'm worried about a drop off where there's something going on that isn't expected and the autonomous vehicle can't figure it out… and I can't figure it out.*" Others, like **P2**, raised concerns about reaction time to "*a fallen tree or a deer in the road.*" Ultimately, participants stressed the fundamental importance of safety in transportation, as **P10** asserted, "*I want to get to my destination, and safety's a big part of that.*" Overall, these concerns were important to users across each segment of the AV trip and help contextualize the importance of information identified in the sequencing tasks during the interview. For example, environmental awareness concerns during the navigation and exiting phase highlight the need for hazard and obstacle detection, whereas concerns with safety when entering speak to the need for correct vehicle confirmations and seat detection. Safety generally was connected with difficulties and challenges experienced during transportation, as provided in the following.

*Difficulties and Challenges*

Participants described a wide range of challenges in current transportation systems, including spatial confusion, unreliable technology, and negative social experiences. Spatial difficulties—particularly around pick-ups and drop-offs—were a major theme. **P1** said, "*It can be really difficult to find the right vehicle. In fact, I made a mistake not long ago and got into the wrong Uber.*" **P12** added, "*Sometimes they'll pull up right in front of the business or sometimes [...] across the street,*" making it hard to identify the correct location. Navigation apps often complicated rather than clarified travel. **P10** shared, "*So often, when I am using Google Maps [...] it's giving me directions as if I can see.*" **P3** called it "*cumbersome*" to juggle multiple apps, especially when one hand is already occupied. These limitations often forced participants to rely on incomplete or confusing information. Participants also spoke about emotional stress, embarrassment, and dependency on others. **P6** noted, "*Sometimes it can be embarrassing to open a door and someone is sitting there,*" and **P5** added, "*Nobody wants to open the door and accidentally sit in somebody's lap.*" These moments contributed to heightened anxiety during entry. Social and systemic challenges were also frequent, especially for those traveling with guide dogs. **P2** stated, "*They're not supposed to refuse me, but I get refusals every day.*" **P8** added, "*They may not always recognize the cane or if they see the dog they'll just keep driving and cancel the ride.*" These ongoing frustrations underscored the need for AV systems that can reduce reliance on unpredictable human interactions. Taken together, challenges and difficulties expressed by participants supported information items ultimately included in our description logic (Section 5), particularly *Finding the Correct Vehicle* and *Where other Passengers are Sitting.*

*Opportunities*

Participants expressed excitement and optimism regarding the potential benefits and opportunities of AVs. They anticipated increased efficiency, availability, reliability, and independence (which were all reoccurring themes) compared to current transportation options. "*It will reduce unpredictability, so I can just get where I need to go,*" stated **P4,** and **P11** shared "*I'm looking forward to the opportunity to just get up and go whenever I want to, which [...] would be amazing!*" highlighting the desire for spontaneous, unrestricted travel. These responses were closely tied to the prospect of greater independence and personal autonomy, as **P1** expressed, "*[AVs] will enable me to go places anytime I want to go and do it safely and independently, that's the main thing.*" **P4** further elaborated on this, saying, "*I'm very excited for independence. And when I say independence [...] it's really the independence of like, I can just get the car when I need it.*" This theme of independence also extended to the hope of personal AV ownership, to which **P5** stated, "*Well, if I have my own autonomous vehicle, that means I can go work wherever I want, whenever I want, which would be pretty sweet*". **P10** also expressed this wish by saying, "*My hope is that one day I can buy a[n] [autonomous] vehicle... I've gone on dates as a blind person, and I can't tell you how many times I wish that I could go pick her up.*" Ultimately, participants described AVs as a pathway to a more inclusive and empowered future, with **P8** concluding, "*I'm thinking big picture. It's going to lead to more gainful employment for people, more ability to take care of your health, more ability to do leisure activity... so overall, a better quality of life.*"

*Adaptations*

Participants also discussed ways in which they have needed to adapt to overcome their current transportation challenges. To improve spatial awareness **P1** explained, "*I like to use the apps to know where I am when I'm in a moving vehicle so that I have some awareness of when I might be arriving*." However, this reliance on technology was not without its complications, as discussed previously, by the need to multitask and manage multiple apps simultaneously. Time management emerged as a popular adaptation, with participants consistently building in buffer time to account for potential delays. **P1** noted, "*I always booked it [a ride] to get me to a place at least a half hour before I was supposed to be there*," illustrating a proactive approach to unpredictable ride timelines. Preemptive planning was also mentioned as a way to navigate unfamiliar environments, as **P9** articulated, "*...I want some information ahead of time so that I can plan how I'm going to hop out and where I'm gonna go from there. I want to have a battle plan, I'm a planner*." Shared rides presented unique social situations where BLV passengers may rely on others, as **P9** acknowledged, "*...if there are other passengers there, you don't want to be fumbling around... [so I] try to ask other people on the ride for assistance or what have you, which certainly happens sometimes on public transit*." Overall, these adaptive behaviors highlight the resourcefulness that is often necessary and the important role of information to overcome transportation hurdles faced by BLV travelers.

*4.2.2 Sequence Results*

*Navigation*

The results of navigation information ordering and presentation from the interviews are presented in Table 3. Participants consistently prioritized *Finding the Correct Vehicle* as the piece of information most desired (and important) to be presented first (mean order = 1.33), favoring an audio presentation with accompanying haptic/vibrational feedback from a device. Subsequent information items, including *Description of Vehicle* (mean order = 2.67), *Avoiding Obstacles/Hazards* (mean order = 2.75), and *Locating the Correct Door* (mean order = 3.58), were ranked after in the sequence and were predominantly preferred in an audio-only format. The most frequent preferred output source for all items was the user's personal device (i.e., their phone).

**Table 3** Preferences for sequence of information presentation for the Navigating to Vehicle phase

| Information Item | Mean Order (M ± SD) | Modality (N, %) | Output Source (N, %) |
|---|---|---|---|
| Finding Correct Vehicle | 1.33 ± 0.65 | Audio + Haptics (7, 58.33) | Phone (7, 58.33) |
| Description of Vehicle | 2.67 ± 1.37 | Audio (9, 90) | Phone (8, 80) |
| Avoiding Obstacles / Hazards | 2.75 ± 1.29 | Audio (7, 70) | Phone (9, 90) |
| Locating the Correct Door | 3.58 ± 1.08 | Audio (4, 44.45) | Phone (4, 44.45) |

The results of the Friedman's test revealed that the ranking scores were significantly different for the information items, $\chi^2(3) = 15.5$, $p = .001$ with a moderate effect size ($W = .43$). Post-hoc pairwise tests revealed statistically significant differences in order between information for *Finding the Correct Vehicle* and each of the other information types (all $p$'s < .05), with *Finding the Correct Vehicle* being preferred first. The order provided by interview participants, starting with information to help find the correct vehicle and ending with locating the correct door, somewhat contradicted the survey data (where participants indicated that they desired information on knowing the correct door first). When asked why they thought this might be, participants mentioned that knowing the correct door was situational: it would be more important to know earlier if it was a relatively full vehicle in a shared ride ($N = 6$ participants) or if they were worried about being embarrassed by going to the wrong door ($N = 4$ participants).

*Entering*

Table 4 presents the participants' sequencing of information related to entering the vehicle and their preferred presentation modalities. Participants generally considered *Where Other Passengers are Sitting* (mean order = 2.58) as the highest priority, followed by how to *Interact with the Vehicle* (mean order = 2.83) and *Seat Placement* (mean order = 3.08). The *Vehicle Interface Location* (mean order = 4.33), *Safety Feature Location* (mean order = 4.58), and *Vehicle Destination* information (mean order = 5.08) were ranked with moderate importance. Lower in importance and thus presented later in the information sequence were *Vehicle Route* (mean order = 6.00), *Vehicle Cleanliness* (mean order = 7.50), and *Forward Direction* information (mean order = 8.17). Across all information items, participants predominantly preferred an audio-only format for presenting the information and selected both information from the vehicle and the phone + vehicle as their preferred output source.

**Table 4** Preferences for sequence of information presentation for the Entering and Orienting phase

| Information Item | Mean Order (M ± SD) | Modality (N, %) | Output Source (N, %) |
|---|---|---|---|
| Where Other Passengers Sitting | 2.58 ± 1.08 | Audio (7, 63.63) | Phone + Vehicle (6, 54.55) |
| Interact with Vehicle | 2.83 ± 1.70 | Audio (8, 66.67) | Vehicle (6, 50) |
| Seat Placement | 3.08 ± 2.43 | Audio (9, 81.81) | Phone + Vehicle (5, 45.45) |
| Vehicle Interface Location | 4.33 ± 2.46 | Audio (9, 81.81) | Phone + Vehicle (5, 45.45) |
| Safety Feature Location | 4.58 ± 2.27 | Audio (10, 83.33) | Vehicle (6, 50) |
| Vehicle Destination | 5.08 ± 2.35 | Audio (12, 100) | Phone + Vehicle (5, 41.67) |
| Vehicle Route | 6.00 ± 2.22 | Audio (10, 90.9) | Phone (5, 45.45) |
| Vehicle Cleanliness | 7.50 ± 3.21 | Audio (5, 83.33) | Vehicle (3, 50) |
| Forward Direction | 8.17 ± 2.62 | Audio (3, 50) | Vehicle (3, 50) |

Again, the ranking scores were significantly different for the information items, $\chi^2(8) = 37.5$, $p < .001$ with a moderate effect size ($W = .39$). Post-hoc pairwise tests revealed statistically significant differences in rank between *Where Other Passengers are Sitting* and *Forward Direction* ($p = .04$). These results suggest a high concern with seat availability and that finding the correct seat is top of mind for BLV users when entering AVs. However, when told that the prior survey results indicated a desire for knowing the vehicle's destination first, interview participants agreed that this was important, particularly to increase comfort and assuage concerns stemming from prior experiences with getting on the wrong bus or rideshare vehicle. Participants specifically mentioned that knowing where the vehicle was going was important for confirming they were in the correct vehicle ($N = 7$ participants), knowing what to expect and preparedness ($N = 5$ participants), and seeking a sense of comfort ($N = 4$ participants).

*Exiting*

Participants' ordering of information during the vehicle exiting phase, along with their preferred presentation of the information, reveal insights into the group's egress priorities (Table 5). The *Safe Side of Vehicle to Exit* (mean order = 1.75) emerged as the most important piece of information to receive first. Following closely was *Direction + Distance to Destination* information (mean order = 2.58), indicating an interest to maintain spatial awareness post-exit. Information regarding *Vehicle Location* (mean order = 3.25) and knowledge of *Hazards/Obstacles Outside* the vehicle (mean order = 3.33) were also considered relatively important. Comparatively, information about the *Direction and Flow of Traffic* (mean order = 6.33), *Points of Interest* (mean order = 6.50), *Direction Facing* of the vehicle (mean order = 6.50), and other *Passengers Exiting/Entering* (mean order = 6.75) were deemed less essential to include. Consistent with the other trip segments, audio-only presentation was favored across all of the information items. As with the navigation phase, the preferred output source was most commonly the user's phone.

**Table 5** Preferences for sequence of information presentation for the Exiting and Orienting phase

| Information Item | Mean Order (M ± SD) | Modality (N, %) | Output Source (N, %) |
| --- | --- | --- | --- |
| Safe Side of Vehicle to Exit | 1.75 ± 0.75 | Audio (7, 58.33) | Vehicle (6, 50) |
| Direction + Distance to Destination | 2.58 ± 1.44 | Audio (9, 81.82) | Phone (5, 45.45) |
| Vehicle Location | 3.25 ± 2.70 | Audio (9, 90) | Phone (4, 40) <br> Phone + Vehicle (4, 40) |
| Hazards / Obstacles Outside | 3.33 ± 1.44 | Audio (9, 81.82) | Phone + Vehicle (5, 45.45) |
| Direction and Flow of Traffic | 6.33 ± 2.35 | Audio (4, 57.14) | Phone (4, 57.14) |
| Points of Interest | 6.50 ± 2.28 | Audio (8, 100) | Phone (4, 50) |
| Direction Facing | 6.50 ± 2.81 | Audio (5, 71.43) | Phone + Vehicle (4, 57.14) |
| Passengers Exiting / Entering | 6.75 ± 2.22 | Audio (4, 57.14) | Phone (4, 57.14) |

The Friedman's test revealed that ranking scores were significantly different for the information items, $\chi^2(7) = 45.4$, $p < .001$ with a large effect size ($W = .54$). Post-hoc tests demonstrated statistically significant differences in rank between *Safe Side of Vehicle to Exit* and *Direction and Flow of Traffic* ($p = .001$), *Points of Interest* ($p = .002$), *Direction Facing* ($p = .02$), and *Passengers Exiting/Entering* ($p = .002$). Results here were more aligned with the survey than in the previous two phases – participants prioritized information about which side to exit the vehicle in both studies. Whereas hazard identification was selected more frequently to be presented first in the survey, it was still ranked highly in the interviews and not statistically significantly different from the items rank ordered above it (*Vehicle Location*, *Direction + Distance to Destination,* and *Safe Side of Vehicle to Exit*). Indeed, when asked why participants from the survey may have selected hazards/obstacle avoidance first, participants mentioned it was very important, noting a general need for safety and to avoid falling risks ($N = 10$ participants), knowing what to expect outside the vehicle ($N = 6$ participants), and how this information could assist navigation to the destination ($N = 4$ participants).

*4.2.3 Output Source*

When considering how the information sequences identified above should be presented, participants were asked if information should come from the vehicle (e.g., a speaker), their personal device (e.g., a smartphone), another device, or a combination. We report these data for each information item individually in Tables 3-5 but offer combined data in the following. For the navigation phase, personal devices were prioritized (68.30% of all responses), followed by a combination (24.39%), and the vehicle (7.32%). During the entry and orientation phase, a combination of personal device and vehicle was most preferred (38.04%), followed by just the vehicle (36.96%) and just the personal device (25.00%). Finally, during the exiting phase, personal devices were again prioritized (38.36%), followed by a combination (35.62%) and just the vehicle (26.03%). Collapsing these responses together, across all three phases, participants prioritized receiving information from their personal device (43.8% of all responses) instead of from a combination (32.55%) or the vehicle alone (23.30%). As explored more thoroughly in our guidelines section, this preference for information from personal devices speaks to participants being comfortable with their personal devices, the privacy they afford, as well as the accessibility settings to which they are already accustomed.

## 5 Description Logic

Our goal in presenting the following description logic (Figure 7) is to provide a structured source of information that can support BLV users who, as our results indicated, (1) are excited to use AVs, (2) experience challenges with the entry, navigation, and exiting portion of the trip, and (3) are accustomed to using information as an adaptation, preferring audio interaction. Participants also highlighted how cumbersome it can be to have to learn multiple apps, each with their own information and interaction style, which is a consistent finding in the BLV AV research [20]. By presenting a unified structure for natural language audio descriptions tied to tasks known to be difficult for BLV travelers, we offer a practical approach for AV designers to adopt a consistent flow of information from one phase of travel to the next.

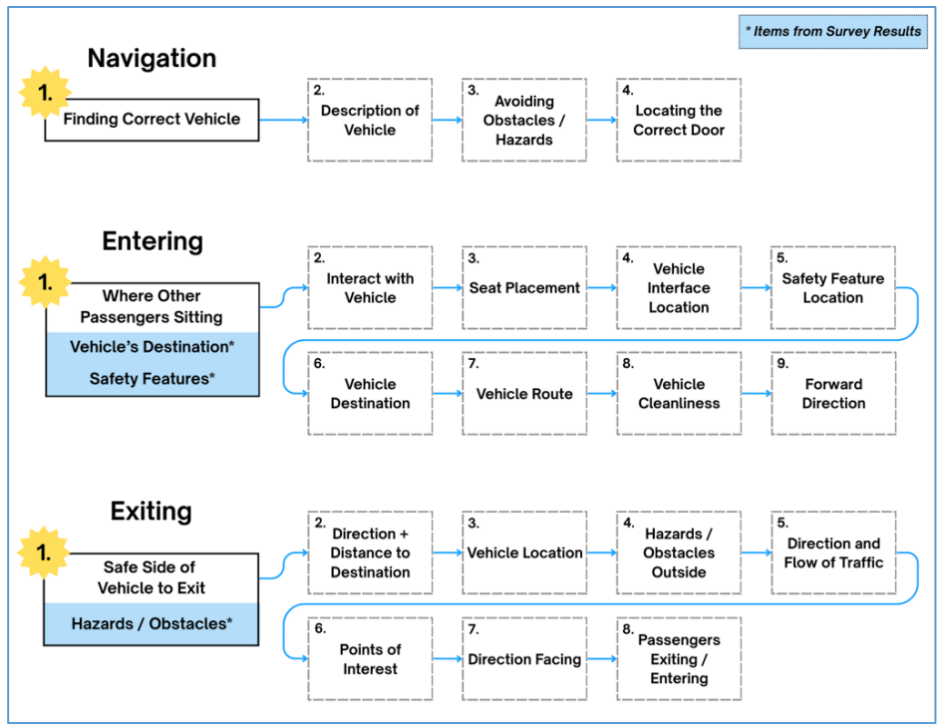

**Fig. 7** Culminating description logic based on Study 1 Survey and Study 2 Interview results

As noted in Section 3 and Section 4, the Study 1 Survey and Study 2 Interview illustrated notable differences in which items should be presented first to BLV users. Opposed to weighting one of these sets of results more heavily than the other, we opted to include the survey results that were statistically significantly more likely to be presented first (i.e., *Vehicle's Destination* and *Safety Features* in the Entering phase and *Hazards/Obstacle Detection* in the Exiting phase) as potential first-items in the final description logic. The logic here, as indicated by our interviews, is that much of the information is situational and should be elevated based on the context of travel. For example, in a crowded pick-up location, confirming the *Vehicle's Destination* can be imperative during the Entry phase to ensure that someone is not entering the wrong vehicle. Likewise, when there are hazards or obstacles to avoid in the destination environment, users want to know about these *Hazards / Obstacles* first. By combining the interview and survey data in this way, we offer a structured approach that also sheds light on how descriptions to support future accessible use of AVs need to be adaptive to the specific context of travel.

## 6 Guidelines and Discussion

Findings from the studies presented here highlight the important challenges and opportunities for BLV users during AV travel, as well as the ways in which information can serve as an adaptation during multiple stages of the trip. For instance, results from both Study 1 Survey and Study 2 Interviews suggest that BLV users face spatial challenges like finding the correct vehicle and navigating pick-up and drop-off locations, as well as concerns with safety when entering and exiting. Moreover, we identify the range of information items that are important to communicate to BLV users to overcome these challenges (e.g., providing information to find the correct vehicle and which side of the vehicle to exit) combined with how these sets of information should be ordered in an accessible interface. The description logic developed as a result can be used by AV designers to develop structured sequences on which contextually relevant information is scaffolded. Although applying the description logic is not the focus of the current work, we envision it serving as a set of variables that AV-related apps can populate with information specific to each trip. Our results show that when doing so, most information should be presented from the user's personal device using audio as the primary information modality. In the following, we explore these findings in conversation with the literature and offer ideas for future work pursuing accessible AVs more generally.

## 6.1 Prioritizing Safety-critical Information Across the Trip

Both the Study 1 survey data and Study 2 interview data indicate the critical role of using information to promote safety for BLV users in the navigating, entry, and exiting phases of AV travel. Participants prioritized safety as a theme during the interviews, as well as in the information order in our description logic, which coincides with existing BLV AV research where safety has been a dominant concern [2, 7, 9]. Results from the Study 1 survey highlight that during the navigation phase, safety concerns emerge due to challenges with avoiding obstacles on the way to the vehicle, as well as finding the correct vehicle. These results echo existing work focused on BLV navigation to AVs that uses computer vision to help users identify objects and localize vehicles and door handles [20]. We add to the existing work by identifying the order of information that should be prioritized, starting first with information to help find the correct vehicle (e.g., distance and direction information), transitioning to obstacle avoidance, and ending with information to find the correct door (e.g., the rear passenger-side door). While participants in the present work did not indicate that finding the correct door and handle was as challenging as indicated by previous work [20], this may have been due to how we combined the tasks of finding the door and finding the handle into a single information item, a potential limitation of our question structure. It is worth noting that as flush door handles become more prevalent, the task of finding the door handle may become increasingly challenging for BLV users. Indeed, designers of future AVs should be cognizant of how design choices can exacerbate current problems for BLV users and, as demonstrated by our interviews, how information can serve as an adaptation to challenges during transportation.

During the entry phase of the trip, safety emerged as a critical theme with participants prioritizing knowing about the safety features available in the vehicle. Interview responses suggest that it is important for AVs to confirm to BLV users that it is safe to get in and to confirm that the user is in the right vehicle and orienting to the correct seat. Interestingly, unlike the navigation and exiting phase, there is an apparent opportunity to explore using the vehicle itself as an output source instead of the user's personal device during the entry phase, as explored more thoroughly in the following subsection. Given these results, designers should ensure that the vehicle's destination is communicated upon passenger entry and that safety features (e.g., hand rails and emergency exits) are appropriately communicated to BLV riders.

Finally, during the exiting phase of the trip, hazard avoidance and communicating which side of the vehicle is safe to exit were prioritized by participants. These data coincide with existing work on exiting AVs by Meinhardt et al. (2025), which emphasized the role of new multisensory interfaces leveraging haptics to communicate both static and dynamic obstacle detection for BLV users [30]. However, our results suggest that users likely prefer using their existing devices primarily via audio, as opposed to new devices leveraging haptics or other multisensory interactions. We explore this finding in the following subsections, offering caveats to our results and suggesting future work.

## 6.2 Designing for BLV Users' Personal Devices

A key takeaway from the two studies presented here is that that BLV users generally prefer to have the majority of AV information that they receive delivered from their personal smartphone. This result coincides with existing accessibility work with AVs [20], with the principal reason being that users often have different preferences and accessibility needs for how information is presented in the UI. As current smart devices have significant universal design (UD) features built-in to the interface for both input and output operations, BLV users are empowered to customize their device with important personalized features like the speech rate of audio output [17]. These accessibility features are native to the device's operating system and deeply embedded into the UI, meaning they work across system states, applications, and usage scenarios, which reduces the learning curve and increases user confidence. This level of access is why an estimated 90% of BLV persons use smartphones, with the vast majority preferring iOS based devices given their incorporation of so many UD design principles in the native UI [36]. The takeaway, as one of our participants aptly put it, is that "*every car is going* *to be different, but [my] phone remains the same.*" Additionally, having the information coming from the smartphone allows users to "*tailor the app or whatever service [they're] using within the phone to tell [them] the information as [they] want it*" (supporting customization). The variety exhibited in our results with respect to which piece of information participants would like to receive first during the navigation phase highlights this need for the user to be able to customize the presentation of the provided information. We argue that AV developers adopting this design decision will not only benefit their userbase, but also their manufacturers, as they simply need to follow well-established accessibility design conventions when creating apps. The alternative (i.e., attempting to build in this level of accessibility into the vehicle control system) will be nontrivial, expensive, and require significant usability testing to avoid conflicts and introduce the potential for broken UI elements upon every update. That said, our data do show that BLV users may prefer some information directly delivered from the vehicle, particularly during the entry phase. How to interact with the vehicle and the location of safety features were items that specifically could be delivered via in-vehicle speakers, whereas others (e.g., seat placement and where other passengers are sitting) could be delivered by both a user's smartphone and the vehicle itself. This finding supports approaches in the literature using in-cabin audio to support BLV users during the trip [19]. We contend that in these combined instances, there may be an opportunity to explore multisensory information (e.g., haptics on the phone and audio from the vehicle) opposed to audio only, as discussed in the following.

### 6.3 Emphasizing Audio Based Interaction?

Our results show a clear preference for BLV users receiving information from AVs using audio as the primary interaction modality. This finding coincides with existing research by Fink et al. (2023) [18] for BLV users across visual status subsamples (mild visual impairment, moderate, and legally blind). However, as recognized in this previous work, BLV users tend to have more experience with audio-based interaction than other interaction modalities (e.g., haptics), meaning self-reported preferences available in the literature may be impacted by experiential or cognitive bias. Indeed, when BLV participants have been exposed to multisensory UIs leveraging haptic interactions in a transportation context, results have been on par with or even outperformed audio-only approaches [22, 29, 30]. This apparent contradiction between self-reported survey and interview results with experimental device testing suggests that more work is needed to expose BLV users to multisensory spatial applications. We argue that as multisensory UIs proliferate, designers should consider how to integrate audio with other modalities like haptics, particularly when a user's personal device can be used (as has been done in recent work with vibro-audio interactions on smartphones [22, 32]).

### 6.4 Limitations

Due to the hypothetical nature of this study, the BLV participants' answers may be skewed toward their current lived experience with human-operated rideshares (e.g., Uber and Lyft). As a result, their answers are based on their current knowledge and experiences with technology and are not necessarily representative of a hypothetical (imagined) perspective in which non-human-operated AV rideshares exist. As this technology slowly migrates from the hypothetical to the practical, further study and analysis would help to better extrapolate the needs of the BLV community regarding independent AV travel. We recommend follow-up behavioral studies to further explore the ideas and framework outlined in this paper, as these studies may add realism to the hypothetical problems of undeveloped technologies and allow for more concrete observations from both participants and researchers alike.

The qualitative component of our study relied on interviews with a sample size of 12 participants. While this number is consistent with prior work in accessibility research with blind and low vision populations, we acknowledge that it is a reasonably small sample that limits the generalizability of our findings. We recognize that our analysis is more exploratory and focused on identifying key information needs rather than making broad population-level generalizations. Future research should strive to increase sample sizes to enhance the statistical power and external validity of findings within the accessibility research domain.

Our study followed a sequential design, with a survey preceding qualitative interviews. We chose this approach because our goal was to identify a culminating description logic. We opted to hold interviews subsequent to the survey to help contextualize and explain the results via rich qualitative data. Although an alternative approach, such as beginning with interviews, would have also been valid, we believe that the trajectory from a broad survey to more focused interviews was most useful in pursuit of our intended goals and how we used the observed outcomes.

Demographic information analysis was limited in due part to the contract with Qualtrics, who was responsible for recruiting respondents to the survey. As such, we did not collect robust demographic information that could have further stratified the participant groups and analyses. This may be of value for future investigation in order to better analyze if people of certain ages, regions, and backgrounds influence the data. For example, related research has shown that people from the United States, Hong Kong, and China may interpret and react to AVs differently as a result of their cultural background [25]. Our focus, lying primarily on what information was important to BLV travelers in the hypothetical scenarios provided and when this information was best to be delivered, did not necessitate the analysis of this demographic information and as such did not take away from the intent and scope of the paper. Furthermore, we recognize that using a third-party service for participant recruitment is an emerging practice in research with specific trade-offs. While it maximized the likelihood of reaching our target population and obtaining data from a larger sample than is practical to recruit for in-lab studies, this technique did prevent us from engaging directly with participants. To ensure data integrity, we implemented screening questions and carefully reviewed all responses to filter out any that appeared inauthentic, incomplete, or suspicious. However, we acknowledge that the use of such a service introduces a risk of unverified data that is not present in direct recruitment and could be considered a drawback of our study (and any survey research using such services).

## 7  Conclusion

This paper explored the information needs of blind and low vision (BLV) travelers throughout the autonomous vehicle (AV) travel experience (i.e., navigating to, entering, and exiting) and examined the optimal sequencing and presentation of information to support travelers in each of these trip segments. The study utilized a remote survey with 202 BLV respondents to assess the key BLV information needs and priorities when using autonomous transportation. The survey was followed by a slate of qualitative remote interviews with 12 BLV participants to understand current transportation challenges, while also predicting future challenges pertaining to AV travel, and identifying information sequences for addressing these current and future challenges. The resulting *description logic* from the survey and interviews emphasizes the critical importance of context-specific information delivery for AV travel; the information needs of BLV travelers are not uniform across the journey and vary significantly depending on the specific trip segment. Specifically, results highlight that information to find the correct vehicle and about the vehicle itself (e.g., make and model, the correct door to enter) are vital during the navigation phase. Upon entering the AV, the focus shifts to information that builds personal security and safety during the trip, such as knowing where the vehicle is traveling to and where safety features are located. Finally, as BLV travelers prepare to exit the AV, the priority once again centers on hazards, obstacle awareness, and knowing which side of the vehicle to exit to ensure a safe transition from the vehicle to the external environment and their destination. Across each of these stages, users identified the need for audio interaction and prioritized information from their personal device (i.e., their smartphone), while also indicating opportunities to receive information from the vehicle itself (e.g., onboard speakers) during the entry phase. Themes identified within the interviews underscore the current challenges faced by BLV travelers, impacting their independence and safety, but also expressing strong optimism for AVs to transform their mobility. Participants' current adaptations, while resourceful, highlight the limitations of existing transportation systems and provide evidence for the need of more reliable and user-friendly transportation solutions. By elucidating distinct information priorities and sequencing preferences within the context of three individual trip segments, this research contributes a foundational framework for AV technologies/interfaces that afford safer and more efficient wayfinding for BLV travelers. As we look toward the future of autonomous transportation and the opportunities that this technology can provide to users who may benefit the most, defining clear and consistent information standards for accessible use of these vehicles is essential for promoting greater independence and mobility for future BLV travelers.

# APPENDIX

## Appendix A

**Study 1 Survey**

**Screening Criteria: (must meet all three)**

1. At least 18 years old and have a known uncorrected visual impairment
2. Utilize an accessibility device or mobility aid (e.g., screen reader, magnification, white cane, guide dog, etc…)
3. Does not drive – Utilizes transportation such as rideshares, public transit, or private vehicles

**Pre-survey Questions:**

1. How would you classify where you live or where you commute to most often?
    a. Urban
    b. Suburban
    c. Rural

2. What accessibility aid or aids do you use? [check all that apply]
    a. Screen readers such as JAWS and NVDA
    b. Mobile screen readers such as VoiceOver and TalkBack
    c. Magnification
    d. White cane
    e. Guide dog
    f. Braille display
    g. Apps such as Be My Eyes and Aira
    h. Other:_______

3. How would you classify your vision status?
    a. Mild vision impairment
    b. Moderate vision impairment
    c. Severe vision impairment
    d. Blindness

**Instructions:**

Thank you for agreeing to help with our research! You will be presented with 3 scenarios that relate to finding, entering, and exiting an autonomous vehicle. For these scenarios, the autonomous vehicles are able to safely and legally drive themselves and are not accompanied by a human driver or an attendant. Please read through each scenario description carefully, imagine yourself in the scenario, and answer the accompanying questions honestly.

**Identifying and navigating to vehicle:**

*Imagine ordering a ride to the grocery store in an urban setting. An autonomous vehicle (a vehicle that drives itself and does not have a human driver) pulls up close to where you are located on a busy street. You know that it's near the sidewalk and you are standing within 5 meters of it.*

Likert Scale: 1 - 5 (1 = Strongly Disagree, 2 = Somewhat Disagree, 3 = Neither Agree nor Disagree, 4 = Somewhat Agree, 5 = Strongly Agree)

1. It can be challenging to locate the correct door and door handle.
2. It can be challenging to find the correct vehicle.
3. I would like to know information about the vehicle, like the make, model, and year, before I get to it.
4. Avoiding obstacles and hazards that are in my way can be challenging.
5. I am comfortable finding my own way to the vehicle.
6. I would like to know about [blank] before I enter the vehicle. [Typed response]
7. I think that [blank] would best help me navigate to the vehicle.
    a. Honking the vehicle's horn
    b. A wayfinding app that uses spatialized audio and natural language descriptions on my smartphone
    c. A wayfinding app that uses vibrations on my smartphone
    d. Asking someone nearby
    e. Other:_____________

8. Before getting to the vehicle, the *first* thing I want to know is [blank].
    a. The direction of the vehicle
    b. If there is a bike lane
    c. Which door I should go to
    d. Where the car is going
    e. If there is a steep curb or curb cut
    f. If there are other people in the car already
    g. Other:_____________

**Entering and orienting inside vehicle:**

*It's likely that fully autonomous vehicles are going to use a rideshare model where there may be multiple people on board and the seating arrangements may not be like traditional cars. For example, these vehicles could feel more like a bus or a train car where the seats sometimes face each other, run along the side of the vehicle, or could even be moved around. Imagine that you ordered one of these autonomous vehicles to go to the grocery store. It has arrived and you have successfully navigated to the correct door and are about to enter it.*

Likert Scale: 1 - 5 (1 = Strongly Disagree, 2 = Somewhat Disagree, 3 = Neither Agree nor Disagree, 4 = Somewhat Agree, 5 = Strongly Agree)

1. I want to know which direction is forward as I enter the vehicle's interior.
2. I want to know about the location of the control interface (e.g., radio, climate, mapping, etc.) within the vehicle.
3. I want to know about the cleanliness of the vehicle's interior.
4. I want to know what seats are available (i.e., unoccupied) within the vehicle.
5. I want to know about seat placement and the layout of the vehicle's interior.
6. I want to know where safety features (e.g., emergency exits, handrails, etc.) are located within the vehicle.
7. I want to know the route the vehicle is going to take as I enter the vehicle's interior.
8. I want to know about where the vehicle is going as I enter the vehicle's interior.
9. I want to know how to interact with the vehicle (e.g., voice, haptics, gestures, etc.)
10. What else would you like to know as you enter or while you are riding in the vehicle? [Typed response]
11. When entering the vehicle, the *first* thing I want to know is [blank]?.
    a. Where the vehicle is going
    b. The route the vehicle is going to take
    c. Where safety features are located
    d. The seat placement

e. Where other passengers are sitting
    f. The vehicle's cleanliness
    g. Where the control interface is located
    h. How to interact with the vehicle
    i. Which direction is forward
    j. Other:_________

**Exiting and orienting to surroundings:**

*The autonomous vehicle has safely driven you to the grocery store. It has pulled up next to the sidewalk about 20 meters from the store's entrance.*

Likert Scale: 1 - 5 (1 = Strongly Disagree, 2 = Somewhat Disagree, 3 = Neither Agree nor Disagree, 4 = Somewhat Agree, 5 = Strongly Agree)

1. I want to know what direction I'm facing when exiting the vehicle.
2. I want to know where the vehicle is located when I exit the vehicle.
3. I want to know about the direction and flow of traffic when exiting the vehicle.
4. I want to know if there are other passengers trying to exit or enter the vehicle.
5. I want to know what hazards or obstacles may be outside of the vehicle.
6. I want to know what points of interest are in the immediate area.
7. I want to know which side of the vehicle is safest to exit through.
8. I feel comfortable exiting the vehicle on my own without assistance.
9. What else would you like to know before or after you exit the vehicle? [Typed response]
10. When exiting the vehicle, the *first* thing I want to know is [blank]?
    a. Which side of the vehicle I'm exiting
    b. What points of interest are around me
    c. What hazards or obstacles may be outside the vehicle
    d. If there are passengers exiting or entering the vehicle
    e. What the direction and flow of traffic is
    f. Where the vehicle is located
    g. What direction I'm facing
    h. Other:_______

**[End of survey]**

## Appendix B
**Study 2 Pre-Interview Worksheet**

**Summary:** The following worksheet is designed to help engineers decide what order accessible information should be presented in future technology that assists with finding and using autonomous (driverless) vehicles.

**Instructions:** When filling out this worksheet, think about what information would be most important for you to access first and then how information should be ordered after that depending on the scenario. Use numbers to rank items in the following scenarios in the order in which you'd would want them to be presented to you. Feel free to eliminate items that are unimportant to you by ranking them with a zero (0). For example, if you would want to know the car's color first, you would place the number 1 next to [Rank: ] for that item. If you don't care about the vehicle's color, mark that item with the number 0 next to [Rank: ].

**Navigating to the vehicle**

Scenario: Imagine ordering a ride to the grocery store in an urban setting. An autonomous vehicle (a vehicle that drives itself and does not have a human driver) pulls up close to where you are located on a busy street. You know that it's near the sidewalk and you are standing within 20 feet of it.

When thinking about an interface that can give you information on the way to that autonomous vehicle, the following order of information would make the most sense or be preferable to you:

[Instructions: Put the following items in the order that you would like them presented to you. Eliminate items you consider unimportant by ranking them with a 0.]

- Information and/or assistance for finding the correct vehicle (e.g., distance and direction information) [Rank:   ]
- Information and/or assistance for locating the correct door or door handle [Rank:   ]
- Information and/or assistance for avoiding obstacles and hazards when navigating to the vehicle [Rank:   ]
- Information and/or assistance about the vehicle itself (e.g., make, model, color, year) [Rank:   ]

**Entering and Orienting**

Scenario: It's likely that fully autonomous vehicles are going to use a rideshare model where there may be multiple people on board and the seating arrangements may not be like traditional cars. For example, these vehicles could feel more like a bus or a train car where the seats sometimes face each other, run along the side of the vehicle, or could even be moved around. Imagine that you ordered one of these autonomous vehicles to go to the grocery store. It has arrived and you have successfully navigated to the correct door and are about to enter it.

When thinking about an interface that can give you information when entering and orienting within the autonomous vehicle, the following order of information would make the most sense or be preferable to you:

[Instructions: Put the following items in the order that you would like them presented to you. Eliminate items you consider unimportant by ranking them with a 0.]

- Information about where the vehicle is going [Rank:   ]
- Information about the route the vehicle is going to take [Rank:   ]
- Information about where safety features are located in the vehicle [Rank:   ]
- Information about the seat placement and orientation [Rank:   ]
- Information about where other passengers are sitting [Rank:   ]
- Information about the vehicle's cleanliness [Rank:   ]
- Information about where the vehicle control interface is located [Rank:   ]
- Information about how to interact with the vehicle [Rank:   ]
- Information about which direction is forward [Rank:   ]

**Exiting and Orienting to the Surroundings**

Scenario: The autonomous vehicle has safely driven you to the grocery store. It has pulled up next to the sidewalk about 20 meters from the store's entrance.

When thinking about an interface that can give you information as you exit an autonomous vehicle and orient to your surroundings, the following order of information would make the most sense or be preferable to you:

[Instructions: Put the following items in the order that you would like them presented to you. Eliminate items you consider unimportant by ranking them with a 0.]

- Information about which side of the vehicle is safe to exit [Rank:   ]

- Information about points of interest that are around you (e.g., direction and distance to your destination) [Rank:   ]
- Information about hazards or obstacles that may be outside the vehicle [Rank:   ]
- Information about other passengers exiting or entering the vehicle [Rank:   ]
- Information about the direction and flow of traffic [Rank:   ]
- Information about where the vehicle is located [Rank:   ]
- Information about what direction you are facing [Rank:   ]

This concludes the worksheet, thanks for your input!

**Appendix C**
**Interview Guide**

Hi there, I'm [name]. Thanks so much for participating in this interview being conducted by [anonymized]. The purpose of the interview is to help the design of future transportation that is more accessible and inclusive. To do so, we'll be asking some questions related to your current experience traveling and what you expect might be helpful in future scenarios with autonomous vehicles. We'll also be audio recording the interview to help with our analysis later on.

You've already provided us some information on the worksheet you completed before this interview. Throughout the interview, we'll ask several times for you to imagine traveling to, entering, and exiting a fully autonomous vehicle. By this we mean a vehicle that can safely, efficiently, and legally drive without a human "at the wheel." What questions can I answer before we get started?

**Demographics and Experience**

To begin, we'll go over some demographic information and then talk about your experiences with technology and transportation.

1. Can you state your name, age, and gender identity?
2. Can you explain what your current day to day transportation experience is like?
3. What challenges or difficulties do you face during transportation?
4. How does the mode of transportation (e.g., public busses vs. rideshare) impact your experience?
5. Can you describe your vision loss? This could include the extent, any metrics you're aware of like acuity or light perception, as well as the etiology, or cause, and onset if known.
    a. How do you use your vision, if at all, during your day to day transportation experience?
6. Do you use a mobility aid (cane or dog, magnification device, etc.)
7. Do you use navigation assistance technologies or apps? For example, Blind Square, SeeingAI, AIRA, Good Maps, etc.?
8. So here we're going to ask you to think about fully autonomous vehicles, like we described before, assuming they can drive safely, efficiently, and legally without a human driver. You can think of it like an Uber or Lyft but without a driver at the wheel. What are your overall thoughts on the rollout of these vehicles?
    a. What are you excited about?
    b. What are you worried about?

**Worksheet**

We sent home a worksheet for you to fill out and we have a few questions about your responses.

*Navigation*

The first scenario had you imagine navigating to an autonomous vehicle without a human driver available to provide assistance.

1. Can you briefly talk me through your thought process as you filled out the worksheet? How did you decide on the order?

Great, thanks. Now we're going to read each item you included, in the order that you included them. For each item, we have a few questions prepared. This might feel repetitive but your input is really important for designing future applications. In the interest of time, I'll ask that your responses are brief, just a few words.

[For each item included on worksheet]

The first piece of information you included was:

1. When or how far from the vehicle would you want this information?
2. Would you want that information in every situation or only in certain situations (for example at night)? What situations?
3. How do you imagine that information being best presented, for example through audio, haptics (i.e., touch/vibration), combinations of both?
4. Can you provide an example of what it might sound or feel like?
5. Where would you like the information to come from? Your phone? The vehicle? Another device?

You decided not to include:

Can you explain this decision?

In a survey of people who are blind and low vision we conducted prior to the interviews, the majority of respondents said knowing which door of the vehicle to go to was a priority to know first. Why do you think this was indicated, and under what scenarios do you think it's most relevant?

*Entry/Orientation*

The second scenario on the worksheet had you imagine entering and orienting within an autonomous vehicle without a human driver available to provide assistance.

1. Can you briefly talk me through your thought process as you filled out the worksheet? How did you decide on the order?

Great, thanks. Now, just like before, we're going to read each item you included, in the order that you included them. For each item, we have a few questions prepared. In the interest of time, I'll ask that your responses are brief, just a few words

[For each item included on worksheet]

The first piece of information you included was:

1. When would you want this information? For example, as you're entering, before entering, or once you're inside?
2. Would you want that information in every situation or only in certain situations (for example at night)? What situations?
3. How do you imagine that information being presented the best, for example through audio, haptics, combinations of both?
4. Can you provide an example of what it might sound or feel like?
5. Where would you like the information to come from? Your phone? The vehicle? Another device?

You decided not to include:

Can you explain this decision?

In that survey we mentioned, the majority of respondents said knowing where the vehicle is going was a priority to know first. Why do you think this was indicated, and under what scenarios do you think it's most relevant?

*Exiting/Orientation*

The third scenario on the worksheet had you imagine exiting an autonomous vehicle and orienting to your surroundings without a human driver available to provide assistance.

1. Can you talk me through your thought process as you filled out the worksheet? How did you decide on the order?

Great, thanks. Now, again, we're going to read each item you included, in the order that you included them. For each item, we have a few questions prepared. In the interest of time, I'll ask that your responses are brief, just a few words

[For each item included on worksheet]

The first piece of information you included was:

1. When would you want this information? For example, as you're exiting, before you exit, or once you're outside?
2. Would you want that information in every situation or only in certain situations (for example at night)? What situations?
3. How do you imagine that information being presented the best, for example through audio, haptics, combinations of both?
4. Can you provide an example of what it might sound or feel like?
5. Where would you like the information to come from? Your phone? The vehicle? Another device?

You decided not to include:

Can you explain this decision?

In that survey we mentioned, the majority of respondents said knowing what hazards or obstacles are outside of the vehicle was a priority to know first. Why do you think this was indicated, and under what scenarios do you think it's most relevant?

**Conclusion**

So to wrap up, we have just a few general questions. In that survey I mentioned, respondents brought up safety as a key concern. What safety concerns do you have?

- Do you think it being a new technology is a contributing factor?

In the example sentences that you gave, we're particularly interested in what we call the reference frame of directions. Do you prefer clockface positions, left-right, near-side vs far-side?

Finally, what problems do you think do you think autonomous vehicles will solve with your current transportation experience?

Thank you so much for participating!